\documentclass{edm_article}

\usepackage{multirow}
\usepackage{tabularx}
\usepackage{array}
\usepackage{subcaption}
\usepackage{xcolor}
\usepackage{enumitem}

\begin{document}

\title{Characterizing Students’ LLM Usage Behaviors and Their Association with Learning in Critical Thinking Tasks}

\numberofauthors{3}

%
%
%
%

\author{
%
%
\alignauthor
Minju Park\\
       \affaddr{University of British Columbia}\\
       \affaddr{Vancouver, BC, Canada}\\
       \email{minju.park@ubc.ca}
\alignauthor
Ivan Orozco Vasquez\\
       \affaddr{University of British Columbia}\\
       \affaddr{Vancouver, BC, Canada}\\
       \email{ivanorozco9@gmail.com}
\alignauthor
Cristina Conati\\
       \affaddr{University of British Columbia}\\
       \affaddr{Vancouver, BC, Canada}\\
       \email{conati@cs.ubc.ca}
}

\maketitle

\begin{abstract}
Large language models (LLMs) are becoming increasingly embedded in students' learning practices, yet much of what is known about how students use LLMs and how this usage impacts learning comes from problem-solving domains or constrained experimental settings. We present an analysis of data on LLM usage collected during two offerings of a research-oriented course where students learn to read, reason about, and critique academic papers. Without restrictions on whether or how to use LLMs, students reported their LLM usage practices when asked to do these activities as a series of homework assignments during the course. This paper extends prior work done on data from a single offering of the same course by presenting a refined bottom-up categorization of LLM usage types, cross-labeled by the extent of student initiative these usages entail. Furthermore, we examine how LLM use impacts student learning, measured by performance on three midterms, looking at factors such as frequency and type of usage.
\end{abstract}


\section{Introduction}
\vspace{5pt}
As large language models (LLMs) become more powerful, sophisticated, and accessible, the educational community continues to grapple with understanding what roles these models can and should play in education~\cite{vadaparty2024cs1, macneil2023experiences, lau2023ban}. While some argue that integrating LLMs into coursework offers benefits such as immediate and personalized feedback~\cite{vadaparty2024cs1, macneil2023experiences}, there are widespread concerns that using LLMs may undermine students' learning, prompting calls to prohibit its use until we gain a better understanding of how it can be effectively leveraged (e.g., ~\cite{lau2023ban}). This paper contributes to a growing body of research addressing the critical need to understand how students use LLMs in practice and how such use may support or hinder learning.

Much of the existing research on student usage of LLMs has focused on problem-solving domains, particularly programming~\cite{vadaparty2024cs1, macneil2023experiences, rasnayaka2024empirical, bull2023generative, kazemitabaar2023novices, ghimire2024coding, brender2024s}. These domains tend to emphasize solution-seeking and correctness, which has led prior work to frame LLM use primarily in terms of debugging or generating solutions and answers. In contrast, many general academic skills---such as reading, reasoning, and critical thinking---require interpretation, judgment, and the construction of understanding rather than arriving at a single correct solution. In such contexts, LLMs may shape how students think and engage with material in more nuanced ways that are not well captured by studies centered on problem-solving tasks.

Another limitation of prior work on students' use of LLMs is that many studies rely on controlled experiments or structured interventions, where students are instructed to use LLMs in predefined or constrained ways. Although some work has examined LLM use in authentic course assignments~\cite{grande2024student}, students were still guided toward specific forms of interaction, which may not reflect how they would naturally choose to use LLMs.

In this paper, we investigate LLM use in a learning context that has received little prior attention: a research-oriented course in which students learn to read, reason about, and critique academic papers. We examine students’ LLM use during two offerings of a real university course by drawing on students' self-reported usage in weekly assignments, which allow us to capture organically occurring LLM use within an in-the-wild classroom setting where students were free to decide whether to use LLMs and how to use it.
The analysis presented in this paper extends prior work on data from a single offering of the same course~\cite{orozco2025emergent}. We adopt a similar bottom-up, data-driven approach which characterizes LLM use based on students' actual reported practices rather than prescribed behaviors or predefined categories, but we present a refined bottom-up categorization of LLM usage types, cross-labeled by the extent of student initiative these usages entail.
We identify patterns in usage types and frequency across assignments, and we examine how LLM use impacts student learning, measured by performance on three midterms.  

Our analysis shows a wide variation in how students used LLMs---differing in frequency, timing, and the types of support they sought. Performance analyses showed that students who used LLMs in their assignments scored lower than non-users on the first midterm, although this gap decreased as the term progressed. Within LLM users, those who used LLMs in more than half of their assignments consistently earned lower average exam scores than those who used them less often. Similarly, students who engaged in more student-driven prompting---providing meaningful input rather than relying on the model to produce most of the content---tended to score higher, though this performance gap also narrowed over the semester. Taken together, our results suggest that, although students may benefit from guidance on how to use LLMs effectively, especially earlier on in the course, given enough time and opportunities to experience the impact of LLM usage on their exam performance, many manage to decide if and how to integrate this usage in their study process without hindering learning. 

In summary, our work contributes to the growing need to understand  whether and how students use LLMs, and how such usage affects learning, by providing insights in a context that is unique in two ways: first, we target the previously unexplored tasks of reading, analyzing and critiquing research papers, which require interpretation and reasoning skills beyond those involved in the most commonly investigated task of programming, problem-solving and essay writing. Second, the data used for our analysis was collected in a real course setting, where students were given complete freedom on if and how to use LLMs for their assignments.
Together, these contributions broaden current understanding of how students are using LLMs in their learning and highlight the importance of investigating LLM use in ecologically valid settings where students' engagement with these tools emerges organically.

\section{Related Work}
\vspace{5pt}
In this section, we review prior work on students' use of LLMs in education along two dimensions: (1) educational contexts of student LLM use, with a focus on differences across domains and instructional settings, and (2) approaches to analyzing students' LLM interactions, including both top-down and bottom-up approaches.

\subsection{Educational Contexts of Student LLM Use}
\vspace{5pt}
Prior research on students' use of LLMs in educational settings has predominantly focused on problem-solving domains, particularly programming and software development. Within this line of work, many studies examine how LLMs can be intentionally integrated into learning activities through structured instructional designs or interventions. For instance, MacNeil et al.~\cite{macneil2023experiences} examine how specific LLM-supported interventions, such as LLM-generated code explanations, influence students' learning processes and outcomes, while Rasnayaka et al.~\cite{rasnayaka2024empirical} investigate the incorporation of LLMs into software development projects within project-based learning contexts.
Other works explore how students interact with and perceive LLM-integrated learning environments in similar problem-solving domains~\cite{kazemitabaar2023novices, sun2024investigating}. Although these studies provide valuable insights into the pedagogical potential of LLMs, they mainly focus on LLM use within designed or structured learning settings.

Complementing this perspective of LLM integration in educational contexts, a smaller set of studies examines how students themselves use LLMs during learning, often by analyzing student prompts or interaction logs. This work has also largely centered on problem-solving domains, where LLM use is closely tied to generating correct answers and solutions~\cite{ma2026examining, sawalha2024analyzing}. As a result, existing empirical characterizations of student-initiated LLM use are largely shaped by problem-solving tasks and contexts.

Beyond problem-solving domains, there is emerging but still limited work examining LLM use in other academic domains. A few studies have begun to explore contexts such as academic writing or discussion-based learning, yet these investigations typically rely on structured forms of LLM use. For example, prior work on academic writing has analyzed student prompts within an LLM-embedded system designed to scaffold writing tasks~\cite{kim2025students}, while studies in ethics education have examined discussion activities in which students were guided to use LLMs in prescribed ways~\cite{grande2024student}. Taken together, these studies illustrate early efforts to extend LLM research beyond problem-solving domains, but they also show that existing evidence remains narrow in scope. Because different academic domains require and emphasize different skills, findings derived from problem-solving contexts may not generalize to learning activities centered on other skills. This observation underscores the need for further studies of students’ LLM use across a broader range of domains and academic skills.

\subsection{Approaches to Analyzing Students' LLM Usage}
\vspace{5pt}
Prior studies have employed a range of analytical approaches to examine how students use LLMs, with many relying on top-down categorization schemes. A growing body of work proposes standardized taxonomies or frameworks for analyzing human-LLM interactions or prompt strategies in more general, non-educational contexts~\cite{gao2024taxonomy, santu2023teler}. Related work in prompt engineering has also developed comprehensive categorizations of prompting techniques through reviews and conceptual frameworks~\cite{schulhoff2024prompt, giray2023prompt}. Building on these approaches, some educational studies apply established categorizations to student data to examine patterns of LLM use in coursework, such as analyzing student prompts according to usage categories~\cite{mollick2023assigning}.

While these top-down approaches provide useful analytical structure, they rely on predefined categories and examine how student behavior fits within them. As a result, they may overlook how students actually appropriate LLMs in practice, including uses that do not align neatly with existing taxonomies. Prior work has noted that much research on prompting and LLM interaction focuses on techniques developed or analyzed by researchers rather than on students' naturally occurring behaviors in learning contexts~\cite{sawalha2024analyzing}. This gap highlights the need for complementary approaches grounded in students’ observed LLM use.

Beyond top-down categorization approaches, some studies have adopted inductive or bottom-up methods that ground analysis in students’ actual LLM interactions. Jelson et al.\cite{jelson2025empirical}, for example, conduct their analysis in the context of academic writing and explicitly anchor their categorization in Flower and Hayes' cognitive model of writing~\cite{flower1981cognitive}, organizing LLM-supported activities around the phases of planning, translating, and reviewing.
Within each phase, they further inductively coded students' prompts based on the specific ways learners engaged with the LLM, allowing the final scheme to emerge from observed behaviors rather than predefined functional categories.
Grande et al.~\cite{grande2024student}, by contrast, take a fully inductive approach, conducting qualitative thematic analysis of LLM-supported ethics discussions to surface recurring themes without constructing an explicit taxonomy.
Sawalha et al.~\cite{sawalha2024analyzing} similarly use thematic analysis to identify prompting strategies in problem-solving tasks. They examine how learners use LLMs to complete quiz-style programming tasks and derive a two-level structure, distinguishing among broad prompting types and the specific strategies within them. Although these strategies vary in complexity, they are all oriented toward obtaining or refining correct answers, reflecting the goal-directed nature of problem solving.

As these examples illustrate, coding schemes naturally reflect the demands of their respective learning domains. Domains that center on interpretation, critique, and reasoning require different forms of support, and in this sense, our work parallels that of Orozco et al.~\cite{orozco2025emergent}.
They developed an emergent coding scheme grounded in students’ interactions from a single research-oriented course, deriving usage types directly from observed behaviors rather than predefined taxonomies. The analysis and results presented in this paper extend this previous work by expanding the dataset to include an additional course offering from a different term. Using the coding scheme from~\cite{orozco2025emergent} as a foundation, we iteratively refined and extended the codes as we analyzed the expanded dataset to capture the full breadth of students' LLM use, yielding a more robust characterization of students' LLM usage patterns.

\section{Data Overview}
\vspace{5pt}
This section provides an overview of the course in which the data were collected and the composition of the dataset used for analysis.

\subsection{Course Structure}
\vspace{5pt}
Our exploratory analysis was conducted in a third-year undergraduate course offered at a public North American institution. The course’s objective was to teach students research methods in the areas of Cognitive Systems, covering the skills and principles common to all forms of research, such as critical thinking and communication. As part of the course, students were assessed using 14 weekly take-home assignments and 3 in-class midterm exams distributed across the term, with the take-home assignments being graded for completion while the in-class midterm exams were graded for quality. The take-home assignments required students to read an assigned research paper on the general topic of human-AI interaction and complete two tasks:

\begin{enumerate}
  \item \textbf{Critical Summary:} Summarize the assigned paper, focusing on the motivations of the work, the proposed solution, how the solution was evaluated, the contributions of the work, and specific aspects related to the details of the human-AI interaction.
  \item \textbf{Discussion Points:} Brainstorm discussion points identifying and discussing weaknesses in the presented research, relating the research to general issues in the field, making connections with other readings, and proposing alternative approaches.
\end{enumerate}

Students were permitted to use LLMs for their take-home assignments, provided they submitted a brief LLM-use report documenting the model(s) used, the prompts, and the generated output, as required by institutional policy. In contrast, the in-class midterms required students to complete the same types of tasks without any assistance.
Each midterm mirrored the assignment format: students read a research paper and produced a critical summary and discussion points, as described above.

In summary, the data used for our analysis was collected as part of regular course activities and institutional requirements, including mandatory LLM-use reporting. Thus, it complies with the university’s policies regarding the ownership and use of course materials for pedagogical quality assurance (Policies 81 and LR11).

\subsection{Dataset Description}
\vspace{5pt}
The data used in this paper was collected across two academic terms of the course: Term~1 included 37 undergraduate students, and Term~2 included 31, for a total of 68 students. Both terms followed the same course structure, assessment design, and LLM usage policies.

The dataset includes students' take-home assignment submissions along with their LLM-use reports (for those who used LLMs).
It also includes scores from the three in-class midterm exams, which serve as evidence of students' unaided performance on tasks structurally similar to the take-home assignments.

In addition, background information about participating students was collected through a brief demographic survey. The survey asked students to report their coding ability (categorized as strong, moderate, or weak), their academic major, and their prior coursework in relevant areas (Human–Computer Interaction, Artificial Intelligence, and Machine Learning). For students in the Cognitive Systems program, majors were recorded along with their specific streams, which correspond to different disciplinary specializations (e.g., psychology, linguistics, philosophy, or computer science). Students' broader disciplinary background---such as whether they belonged to Arts or Science---was then inferred from their reported major.
All collected data were anonymized prior to analysis.

\section{Methods}
\vspace{5pt}

\begin{table*}[!t]
\centering
\renewcommand{\arraystretch}{1.2}
\begin{tabularx}{\textwidth}{|>{\centering\arraybackslash}p{2.2cm}|p{2.6cm}|X|p{5.5cm}|}
\hline
\textbf{Support Type}
& \multicolumn{1}{c|}{\textbf{Category}}
& \multicolumn{1}{c|}{\textbf{Definition}}
& \multicolumn{1}{c|}{\textbf{Sample Excerpt from LLM-Use Reports}} \\
\hline
\multirow{2}{=}{\centering \\Student-driven Support}
& \textbf{IL}: \textsc{Improving Language}
& Students provide their own draft text and ask the LLM to improve language (e.g., clarity, grammar).
& \textit{``Prompt Used: Here is my summary for the attached paper: [student's draft] Revise it for any grammatical issues.''} \\ \cline{2-4}
& \textbf{CI}: \textsc{Confirming Ideas}
& Students propose their own ideas or drafts and ask the LLM to validate the content or to check whether the draft meets the assignment requirements.
& \textit{``Prompt Used: Does this address all the required points? [student's draft]''} \\ \hline
\multirow{3}{=}{\centering \\LLM-driven Support}
& \textbf{GI}: \textsc{Generating Ideas}
& Students ask the LLM to suggest ideas (e.g., topics, arguments, outlines) for writing.
& \textit{``I used ChatGPT 4.0 to generate ideas for discussion points beyond what I could initially come up with after reading the paper.''} \\ \cline{2-4}
& \textbf{S}: \textsc{Summarizing}
& Students ask the LLM to give an overview of the paper, which is separate from the assignment's more analytic critical summary task.
& \textit{``Prompt Used: Summarize this paper.''} \\ \cline{2-4}
& \textbf{GA}: \textsc{Generating Answers}
& Students ask the LLM to generate an answer or text from scratch (discussion points or critical summary).
& \textit{``In the prompt, I passed the file of the paper and the document of the discussion document [assignment format] and asked it to complete it.''} \\ \hline
\multirow{2}{=}{\centering \\Comprehension Support}
& \textbf{IE}: \textsc{Information Extraction}
& Students ask the LLM to extract specific information from a paper.
& \textit{``Asked GPT to help with finding the mitigating solution against the potential usability threats.'} \\ \cline{2-4}
& \textbf{CC}: \textsc{Clarifying Concepts}
& Students ask the LLM for further explanation to deepen understanding of certain concepts.
& \textit{``Prompt Used: What's the difference between formula 1 and formula 2 in the paper?''} \\ \hline
\end{tabularx}
\caption{Categories of students' LLM usage grouped by support type.}
\label{tab:llm_usage_categories}
\end{table*}

To determine whether data from Term~1 and Term~2 could be analyzed jointly, we compared the distributions of students' backgrounds across the two terms, focusing on self-reported coding ability and disciplinary background. Although the cohorts were not identical, their distributions were similar enough for our purposes. For coding ability, strong or moderate proficiency differed by 6.22\% (83.78\% in Term~1 vs. 90.00\% in Term~2). For disciplinary background, the proportion of Science students differed by 10.55\% (62.16\% vs. 51.61\%), with the remaining percentages falling into the Arts category. Given these broadly comparable patterns, we analyzed the two cohorts together in subsequent analyses.

\subsection{Coding Scheme for LLM Usage}
\vspace{5pt}
A bottom-up coding scheme was used to characterize students’ LLM usage. The analysis began with the coding framework introduced in earlier work~\cite{orozco2025emergent}, which identified seven usage categories: \textit{Improving Language}, \textit{Paragraph from Bullets}, \textit{Clarifying Concepts}, \textit{Confirming Ideas}, \textit{Generating Ideas}, \textit{Summarizing}, and \textit{Generating Answers}. This framework served as an initial guide and was iteratively refined while analyzing the expanded two-term dataset to better capture the full range of usage behaviors observed.

Each submission was coded individually, and multiple categories were assigned whenever a student's prompts reflected more than one form of usage, such as when they engaged in several distinct interactions within the same submission.
Using the existing categorization~\cite{orozco2025emergent} as a starting point, two authors independently coded all LLM-use reports.
Inter-rater reliability, assessed with Krippendorff's alpha, which is appropriate for nominal data with multiple labels per entry, indicated low agreement beyond chance across all categories ($\alpha$ range = -.01 to .03).
This low reliability suggested that several category definitions overlapped or required clearer differentiation. The two authors then met to review the coded data, discuss sources of disagreement, and refine the categorization by clarifying definitions and determining where categories should be merged or further distinguished.
Following this refinement process, the authors iteratively discussed and reconciled remaining disagreements until consensus was reached on all coded instances, resulting in a finalized coding scheme and dataset used for analysis.

The refinement process led to modifications of the existing categories as follows: first, we consolidated ``Paragraph from Bullet Points'' into ``Improving Language.'' In the initial categorization, ``Paragraph from Bullet Points'' captured submissions in which students provided bullet-point inputs and asked the LLM to convert them into a paragraph. However, this process inherently involved improving wording, structure, or clarity, resulting in substantial conceptual overlap with the ``Improving Language'' category. Moreover, a closer examination of these submissions revealed considerable variability in the level of detail provided by students, ranging from brief keyword-style bullet points to full sentences or paragraph-length inputs. This inconsistency, combined with the significant overlap in function, motivated the decision to merge ``Paragraph from Bullet Points'' into the broader ``Improving Language'' category.

Second, refinements to the ``Clarifying Concepts'' category led to the creation of a new category, ``Information Extraction.''
Initially, queries that asked the LLM to locate specific information within the assigned paper were grouped under ``Clarifying Concepts'' since that category broadly covered student questions about course material.
However, closer analysis showed that these queries were not aimed at deepening conceptual understanding but at retrieving specific information from the paper---such as checking whether particular details appeared in the text rather than interpreting the underlying concepts.
This distinction led to the separation of the two categories, where questions that sought conceptual clarification without explicitly requesting information from the paper were retained under ``Clarifying Concepts,'' whereas queries that involved searching for text or data explicitly stated in the paper were classified as ``Information Extraction.''

\begin{figure}[!t]
    \centering
    \includegraphics[width=1\linewidth]{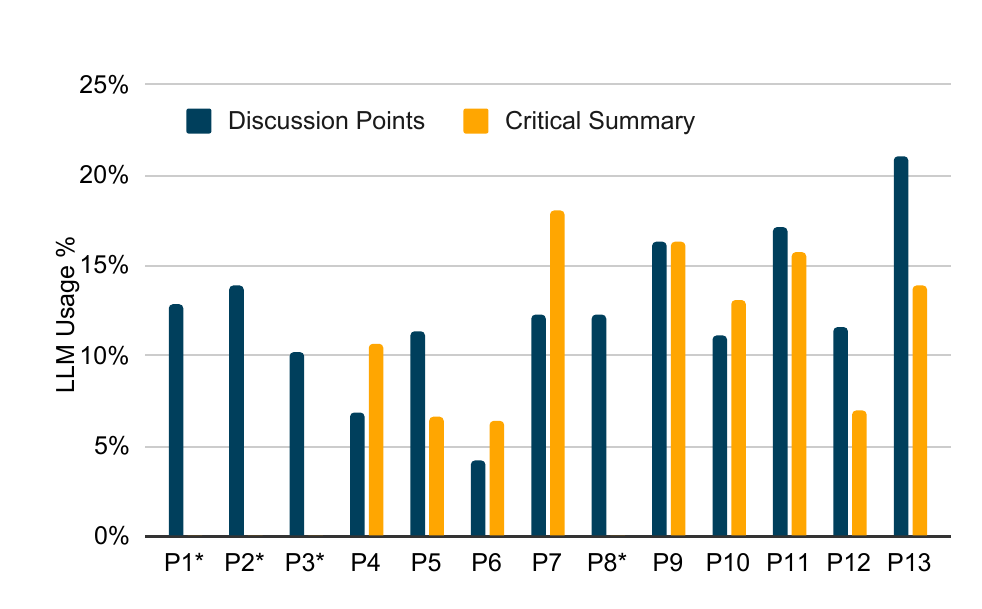}
    \caption{LLM usage rates for discussion points and critical summary tasks per paper. Asterisks indicate assignments with discussion points only.}
    \Description{The figure presents LLM usage rates across papers, with separate bars for discussion point and critical summary tasks. For each paper, the height of each bar represents the proportion of submissions that involved LLM use. Asterisks denote papers where only discussion points were assigned, so no critical summary usage is shown for those cases.}
    \label{fig:usage_trends_per_paper}
\end{figure}

This process resulted in our final set of seven categories---\textsc{Improving Language}, \textsc{Confirming Ideas}, \textsc{Generating Ideas}, \textsc{Summarizing}, \textsc{Generating Answers}, \textsc{Information Extraction}, and \textsc{Clarifying Concepts}.
Definitions of each category and corresponding sample excerpts from the LLM-use reports are presented in Table~\ref{tab:llm_usage_categories}. The table also shows that the categories are organized into three higher-level types of support types based on the role the LLM played in the student's work:
\begin{itemize}[leftmargin=10pt, itemsep=2pt]
    \item \textbf{Student-driven Support}: cases in which students provided initial ideas or drafts and used the LLM for feedback, refinement, or confirmation (includes \textsc{Improving Language} and \textsc{Confirming Ideas})
    \item \textbf{LLM-driven Support}: cases in which students used the LLM to generate content from scratch with minimal prior input (includes \textsc{Generating Ideas}, \textsc{Summarizing}, and \textsc{Generating Answers})
    \item \textbf{Comprehension Support}: uses of LLMs to help understand the assigned paper or related concepts (includes \textsc{Information Extraction} and \textsc{Clarifying Concepts})
\end{itemize}
\vspace{-5pt}

\begin{figure*}[!t]
    \centering
    \includegraphics[width=\textwidth]{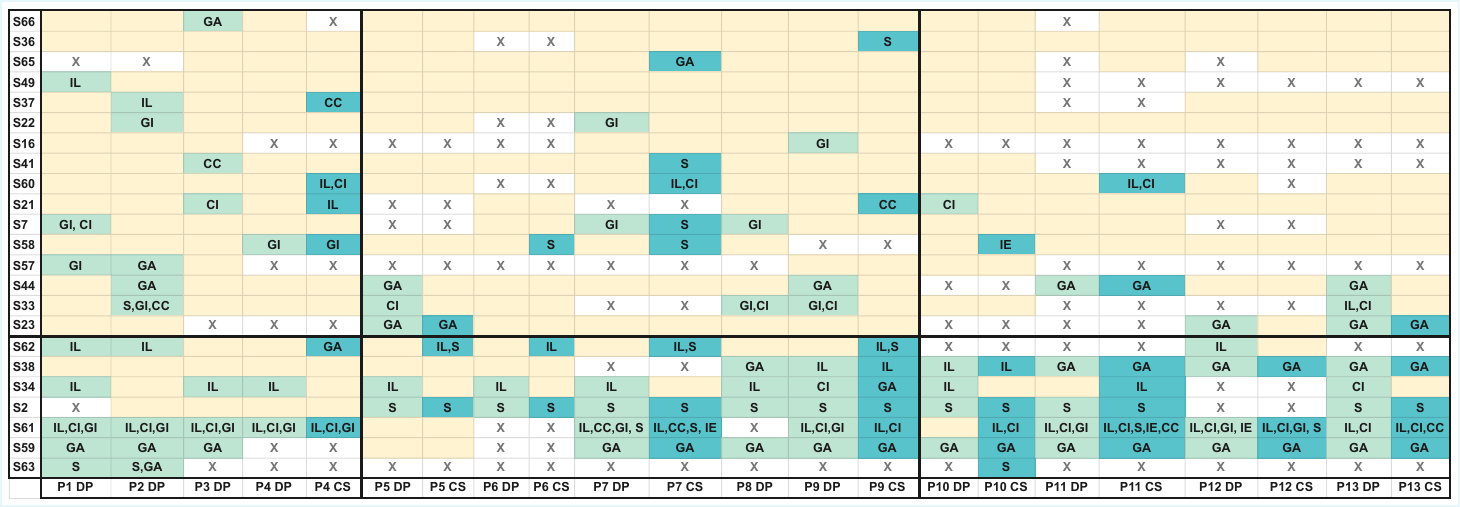}
    \caption{Temporal visualization of reported LLM use across assignments.}
    \Description{A temporal grid of LLM usage where columns represent assignments and rows represent students. Each cell encodes submission status and LLM use: ``X'' denotes no submission, blank cells indicate no LLM use, and labeled cells indicate the specific type(s) of LLM use reported for that assignment.}
    \label{fig:temporal_visualization}
\end{figure*}

\subsection{Analysis Procedure}
\vspace{5pt}
Analyses were conducted at two complementary levels. At the submission level, LLM usage behaviors were examined for each individual take-home assignment, with submissions coded into one or more usage categories based on the coding scheme described above. Category frequencies and usage patterns were then summarized across all submissions.

At the student level, LLM usage data were aggregated across each student's submissions and compared with their in-class midterm exam scores to examine how patterns of LLM use related to learning. This analysis considered both the extent of students' LLM use and the types of support they sought.


We employed descriptive statistics and group-based comparisons to examine overall usage trends, associations between LLM usage and learning, and differences across LLM support types. Analyses focused on identifying patterns of association rather than making causal claims.

\section{Results}
\vspace{5pt}
In this section, we present our findings on how students used LLMs throughout the course and how these usage patterns relate to midterm exam performance, considering both the level of LLM use and differences in LLM usage types.

\subsection{Students' LLM Usage Trends}
\vspace{5pt}

We organize our analysis of students' engagement with LLM into two parts: 
(1) whether and how frequently students used LLMs across assignments over time, and 
(2) how students used LLMs in terms of the functional purposes captured in our coding scheme.

23 out of 68 students (34\%) reported using LLMs in at least one submission.
On average, students used LLMs in 12\% of discussion point submissions and 11\% of critical summary submissions.
Students reported using a range of LLM tools, including ChatGPT (e.g., GPT-3.5, GPT-4, and GPT-4o), NotebookLM, and Claude models.
To see how these rates varied for each assignment, Figure~\ref{fig:usage_trends_per_paper} shows the percentage of students who used LLMs for the two tasks, with papers (P1–P13) arranged chronologically from left to right on the $x$-axis. Note that for some papers, students had to submit only discussion points, no summaries (P1, P2, P3, and P8; indicated with an asterisk in Figure~\ref{fig:usage_trends_per_paper}). The figure illustrates that LLM usage varied across assignments and tasks: for some papers, students used LLMs more for discussion points, whereas for others, usage was higher for the critical summary. A notable trend is the increase in LLM usage for later assignments, particularly for the discussion points task, suggesting that students may have grown more comfortable integrating LLMs into their workflow as the course progressed. While actual usage may be slightly higher due to potential underreporting (e.g., students neglecting to declare LLM use), it is unlikely that the true usage rate differs substantially, as students were explicitly permitted to use AI tools without penalty provided such use was disclosed.

\begin{figure}[t]
    \centering
    \includegraphics[width=\linewidth]{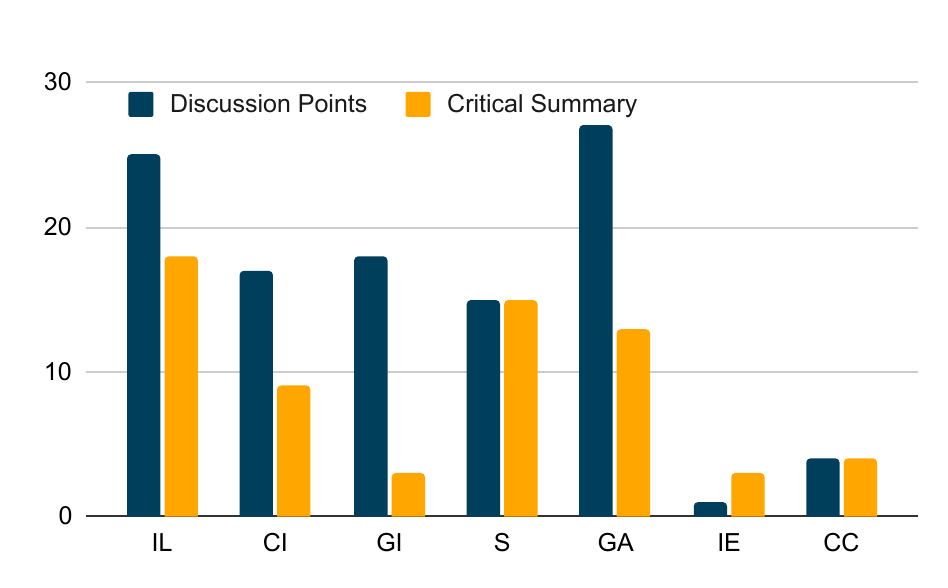}
    \caption{Counts for each LLM usage category for discussion points and critical summary.}
    \Description{Bar chart showing the counts of each LLM usage category, with separate bars for discussion points and critical summary tasks. Each category is displayed along the x-axis, and bar height represents the number of instances coded for that category.}
    \label{fig:code_counts_dp_summary}
\end{figure}

Figure~\ref{fig:temporal_visualization} provides a more detailed view of how each of the 23 students who used LLMs at least once (S$N$ on the $y$-axis) engaged with LLMs for each assignment (P1 through P13 on the $x$-axis).
The papers (P1–P13) are shown in chronological order from left to right. Submissions for each paper (P$N$) are separated in into discussion points (P$N$ DP) and critical summary (P$N$ CS). The types of LLM use categorized for each submission are indicated in each cell by the corresponding code (see Table~\ref{tab:llm_usage_categories}). Cells contain usage codes when the student used an LLM for that task, appear in yellow when the student reported no LLM use, and are crossed out when the assignment was not submitted. Two vertical lines in the middle of the figure mark the timing of Midterm~1 and Midterm~2, and Midterm~3 occurred after the final assignment.
Students are then ordered by their overall frequency of LLM use, with the most frequent users appearing at the bottom and the least frequent at the top.
Because some students did not submit every assignment, usage frequency is computed as the number of assignments in which they used an LLM divided by the number they submitted.
Notably, S63 appears at the very bottom: although their use of LLMs may look sparse visually, this is because they submitted only 3 of the 13 assignments---but used LLMs for all three---resulting in a 100\% usage rate.

The visualization shows a distinct cluster of students who had a much higher usage frequency toward the bottom. Looking at the actual frequencies, a clear breakpoint emerges at S62: students above S62 used LLMs in only 5–33\% of their submissions, while those from S62 downward used them in 50–100\% of their work. This pattern provides a natural division into two groups, with 7 students using LLMs in at least half of their submitted assessments and 16 students whose usage remained below that threshold.
In the rest of this analysis, we will indicate the two groups as \textit{High-Reliance} and \textit{Low-Reliance}, respectively.

These two groups showed different trend of LLM usage over time. Students in the Low-Reliance group showed more concentrated usage early in the term, with an average of three students using LLMs per assignment before Midterm 1. After Midterm 2, this average fell to one, indicating that students in this group not only rely on LLMs only lightly, but they reduced their usage as they grew more comfortable with the assignment tasks over time.
In contrast, within the High-Reliance group, several students (S62, S34, S61, S59) used LLMs consistently throughout the term while others (S2, S38) shifted into heavier use partway through the term. S2 refrained from using LLMs until after Midterm~1, at which point they started using them in every subsequent submissions. Similarly, S38 began to use LLMs at the halfway point of the term, and then continues to use it for every submission until the end of the course.
These varied trajectories within the High-Reliance group suggest that these two students may have adjusted their reliance on LLMs as they grew more familiar with the course structure and their workflow for using them, resulting in increased use later in the term when they felt such usage would no longer hinder their ability to perform the assignment tasks on their own during the midterm.



We next examine how LLMs were used in terms of usage category. 
Figure~\ref{fig:code_counts_dp_summary} presents overall counts for each LLM usage category grouped by assignment tasks (Discussion Points vs Critical Summary).
As the figure shows, \textsc{Summarizing} (S) and \textsc{Clarifying Concepts} (CC) were used equally for the two tasks, which makes sense since both these usages help understand the content of the paper, which is needed for both tasks. \textsc{Information Extraction} (IE) is used more for critical summaries, which also makes sense since writing the critical summary is more directly related to finding specific information in the paper. All the other usages are higher for discussion points, showing that for this activity, which is more open-ended and requires going beyond the content of the paper, students needed more help on \textsc{Improving Language} (IL), \textsc{Confirming Ideas} (CI), \textsc{Generating Ideas} (GI) and \textsc{Generating Answers} (GA).

\begin{figure}[t]
    \centering
    \includegraphics[width=0.8\linewidth]{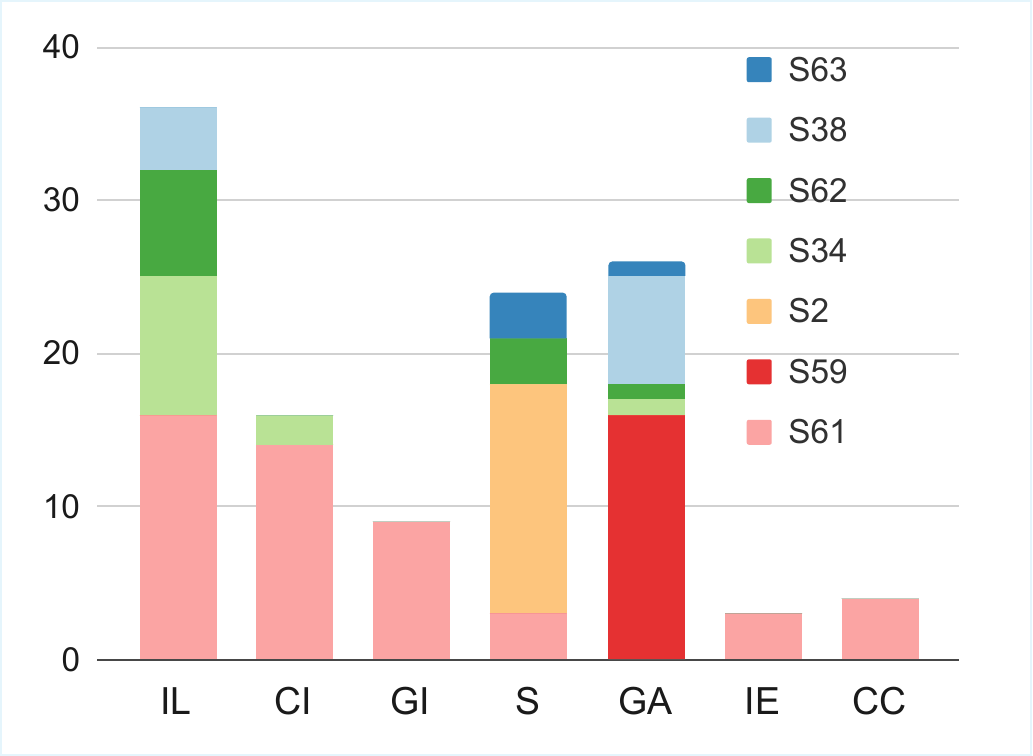}
    \caption{Counts for each LLM usage category, with stacked bars showing the proportion of individual students in the High-Reliance group.}
    \Description{The figure displays the total counts for each LLM usage category, with stacked bars illustrating how these counts are distributed across individual students in the High-Reliance group. Each segment within a bar corresponds to one student's contribution.}
    \label{fig:code_counts_high_reliance}
\end{figure}

\begin{figure}[t]
    \centering
    \includegraphics[width=\linewidth]{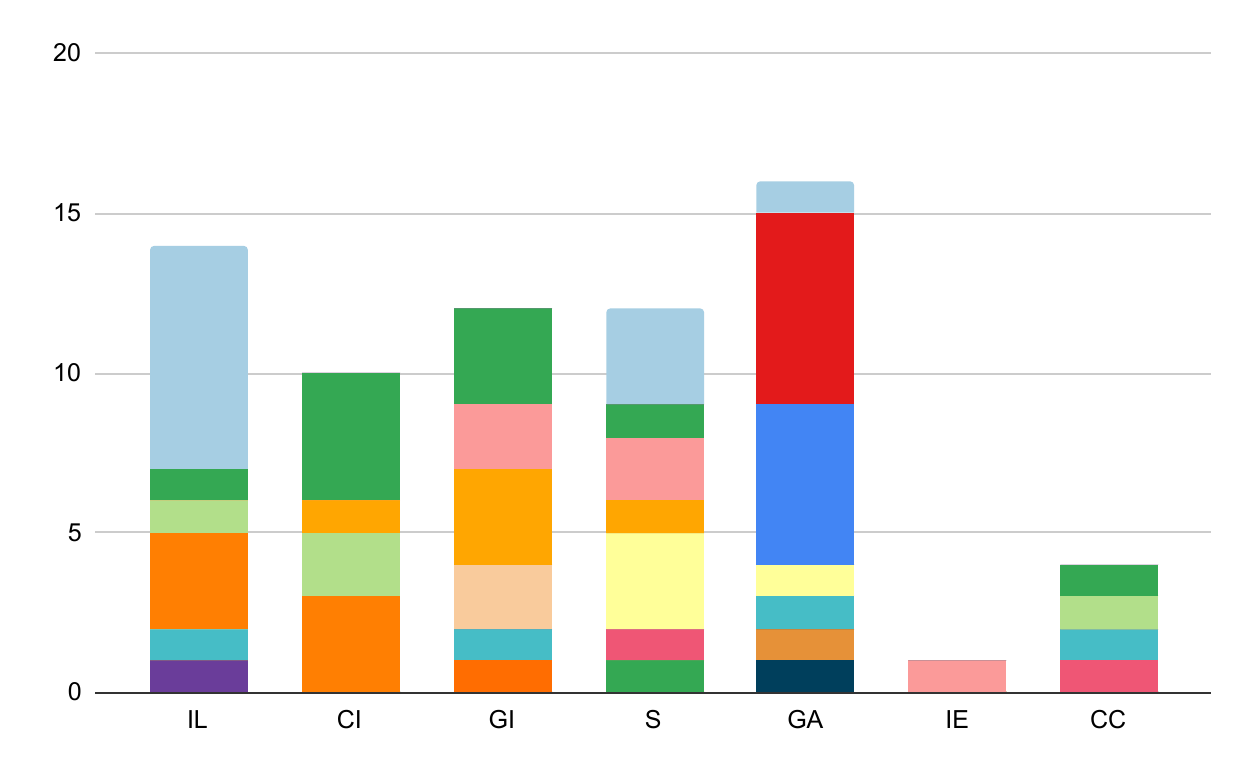}
    \caption{Counts for each LLM usage category, with stacked bars showing the proportion of individual students in the Low-Reliance group. Each color represents a different student.}
    \Description{The figure displays the total counts for each LLM usage category, with stacked bars illustrating how these counts are distributed across individual students in the Low-Reliance group. Each segment within a bar corresponds to one student's contribution.}
    \label{fig:code_counts_low_reliance}
\end{figure}

Figure~\ref{fig:code_counts_high_reliance} shows stacked bar graphs indicating, for each usage category, the proportion of individual students in the High-Reliance group who used it. Figure~\ref{fig:code_counts_low_reliance} shows the same information for students in the Low-Reliance group.

As seen in Figure~\ref{fig:code_counts_high_reliance}, in the High-Reliance group, three categories (\textsc{Generating Ideas} (GI), \textsc{Information Extraction} (IE), and \textsc{Clarifying Concepts} (CC)) were used only by one student, S61. This student used all the categories except for \textsc{Generating Answers} (GA), indicating that, although the student relied heavily on LLMs, they never used them to generate the assignment from scratch. Notably, S61’s midterm scores steadily increased across the term (69\% to 72\% to 79\%), illustrating that extensive LLM use---when focused on forms of support other than full answer generation---can coexist with, and may even accompany, improved performance.
By comparison, S59 used LLMs solely for \textsc{Generating Answers} (GA), and S63 used only \textsc{Generating Answers} (GA) and \textsc{Summarizing} (S), both examples of LLM-driven support.
Finally, S2 used only \textsc{Summarizing} (S), and their performance provides a clear illustration of the potential consequences of this pattern: their midterm scores declined consistently—from 65\% to 48\% to 38\%. After Midterm~1, S2 appears to have increasingly relied on LLM-generated summaries rather than engaging with the full readings, which may have weakened their ability to process new material during the exams.
The remaining students (S62, S38 and S34) spread their LLM usage over both student-driven and LLM-driven categories. 

Figure~\ref{fig:code_counts_low_reliance} shows that, in the Low-Reliance group, each usage type was used by at least four different students aside from \textsc{Information Extraction}, indicating a relatively even distribution of usage across categories. This contrasts with the High-Reliance group, where certain categories are dominated by only a few students.
Notably, even within the Low-Reliance group, \textsc{Generating Answers} (GA) was the most frequently used category. Two students in particular---S44 (red; second segment from the top in the GA bar) and S23 (blue; third segment from the top in the GA bar)---used \textsc{Generating Answers} (GA) more than any other category. As Figure~\ref{fig:temporal_visualization} illustrates, both students tended to rely on \textsc{Generating Answers} (GA) more heavily toward the end of the term.
Interestingly, S44 and S23 showed strikingly similar performance trajectories across the three midterms. Both improved substantially from Midterm~1 to Midterm~2 (S44: 54\% to 82\%; S23: 65\% to 84\%) but experienced a decline on Midterm~3 (S44: 78\%; S23: 76\%). This parallel trend suggests that increased reliance on \textsc{Generating Answers} (GA) later in the course did not necessarily prevent performance gains, but may also not have sustained them through the final assessment.

\subsection{Impact of LLM Usage on Midterm Performance}
\vspace{5pt}

\subsubsection{Performance Differences by Level of LLM Use}
\vspace{3pt}
We examined how students' use of LLMs in take-home assignments impacts their performance in midterm exam performance. For each midterm, we distinguished between students who had used LLMs in any prior assignment and those who had not. Because students adopted LLMs at different times during the term, group sizes varied across midterms. For example, S2 did not use LLMs before Midterm 1 but began doing so between Midterms~1 and~2, and was therefore treated as an LLM user for Midterm 2 and Midterm 3, but not Midterm 1. For Midterm 1, three students were excluded from the analysis: two were absent, and one experienced technical issues at the start of the exam that reduced their available exam time by half. Midterm 2 had one absent student, while Midterm 3 had three students absent.

\begin{figure}[t]
    \centering
    \includegraphics[width=1\linewidth]{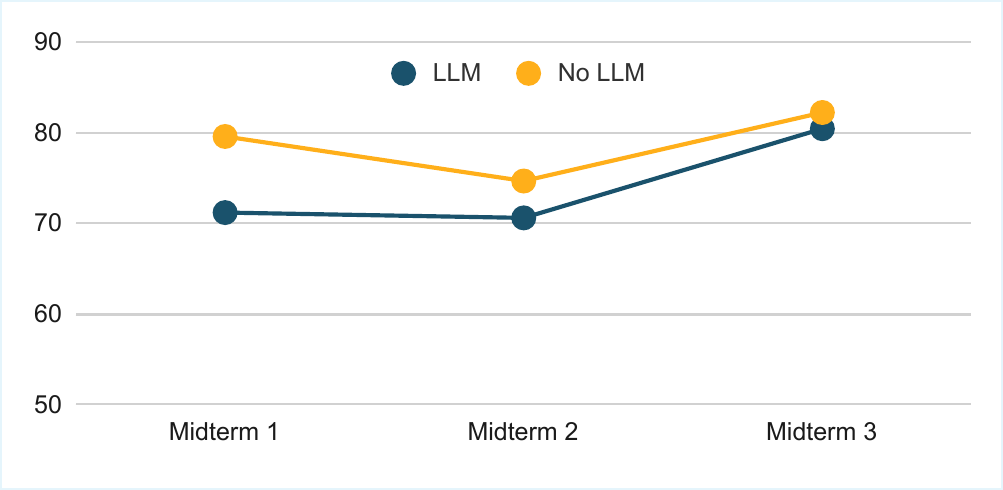}
    \caption{Comparisons of midterm exam scores between students who used LLMs in assignments (\textit{LLM}) and those who did not (\textit{No LLM}).}
    \Description{The figure shows a line graph of midterm exam performance for students who used LLMs and those who did not. The x-axis represents the three midterms, and the y-axis represents average scores. Two lines trace the performance of each group, allowing comparison of trends across exams.}
    \label{fig:LLM_noLLM}
\end{figure}

\newcolumntype{C}[1]{>{\centering\arraybackslash}p{#1}}
\begin{table}[t]
\vspace{5pt}
\centering
\renewcommand{\arraystretch}{1.2}
\begin{tabular}{
    C{0.4cm} | C{1.25cm} | C{0.25cm} | C{0.65cm} C{0.65cm} | 
    C{0.4cm} C{0.4cm} C{0.6cm} C{0.5cm}
}
\hline
\textbf{M} & \textbf{Condition} & \textbf{N} &
\multicolumn{1}{c}{\textbf{M}} & \multicolumn{1}{c|}{\textbf{SD}} &
\multicolumn{1}{c}{\textbf{t}} & \multicolumn{1}{c}{\textbf{df}} &
\multicolumn{1}{c}{\textbf{p}} & \multicolumn{1}{c}{\textbf{d}} \\ \hline

\multirow{2}{*}{\textbf{M1}} 
& No LLM & 44 & 79.60 & 13.15 & \multirow{2}{*}{-2.5} & \multirow{2}{*}{41.8} & \multirow{2}{*}{\textbf{0.016*}} & \multirow{2}{*}{-0.65} \\
& LLM    & 21 & 71.20 & 12.36 &  &  &  &  \\ \hline

\multirow{2}{*}{\textbf{M2}} 
& No LLM & 44 & 74.68 & 11.52 & \multirow{2}{*}{-1.5} & \multirow{2}{*}{52.0} & \multirow{2}{*}{0.133} & \multirow{2}{*}{-0.37} \\
& LLM    & 23 & 70.62 & 9.69  &  &  &  &  \\ \hline

\multirow{2}{*}{\textbf{M3}} 
& No LLM & 42 & 82.24 & 9.04  & \multirow{2}{*}{-0.7} & \multirow{2}{*}{39.6} & \multirow{2}{*}{0.499} & \multirow{2}{*}{-0.19} \\
& LLM    & 23 & 80.46 & 10.60 &  &  &  &  \\ \hline
\end{tabular}
\caption{Comparisons of midterm exam scores between students who used LLMs in coursework (\textit{LLM}) and those who did not (\textit{No LLM}). Here, M denotes Midterm.}
\label{tab:midterm_llm_no_llm}
\end{table}

We first compared midterm exam scores between students who reported using LLMs and those who did not. Figure~\ref{fig:LLM_noLLM} shows line graphs of the averages over the 3 midterms, for the students who did use LLMs (\textit{LLM}) and those who did not (\textit{No LLM}). It shows that students in the No LLM group scored higher on average than the LLM group, although this gap narrowed as the term progressed.
In Midterm 2, this is due to a decrease in performance for the No LLM group, whereas the performance of the LLM group remains similar to Midterm 1. In Midterm 3 both groups improve.
A possible explanation is that, as students became more familiar with the processes entailed in the exam tasks, the impact of LLM usage on midterm performance may have diminished.
Statistical tests and detailed score values are provided in Table~\ref{tab:midterm_llm_no_llm}, showing a statistically significant difference in performance between the LLM and No LLM groups for Midterm 1, with a high effect value.

\begin{figure}[t]
    \centering
    \includegraphics[width=1\linewidth]{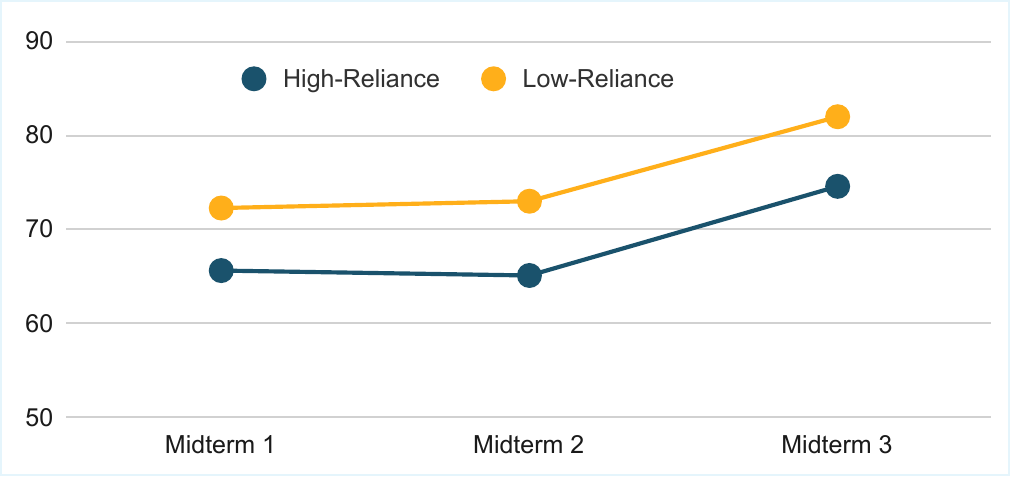}
    \caption{Comparisons of midterm exam scores between students in the High-Reliance and Low-Reliance groups.}
    \Description{The figure shows a line graph of midterm exam performance for High-Reliance and Low-Reliance students. The x-axis represents the three midterms, and the y-axis represents average scores. Two lines trace the performance of each group, allowing comparison of trends across exams.}
    \label{fig:reliance_midterm_linegraph}
\end{figure}

We further investigate the impact of LLM usage on midterm exams by comparing the performance of High-Reliance and Low-Reliance students within the LLM group. Because the group sizes were highly imbalanced (seven vs sixteen), we did not conduct statistical tests; instead, we interpret the descriptive trends shown in Figure~\ref{fig:reliance_midterm_linegraph}, which compares midterm exam scores between the two reliance groups across the three midterms.
On average, Low-Reliance group consistently outperformed the High-Reliance group across all three midterms, with a stable performance gap of approximately 7-8\% throughout the term.
As a matter of fact, the Low-Reliance group performed similarly to the No LLM group, with a performance gap of just 3\%.

\subsubsection{Performance Differences by LLM Usage Types}
\vspace{3pt}
Among students who reported using LLMs, we further examined whether their performances differed depending on the types of LLM support they relied on. Because individual students often engaged in multiple forms of LLM use across assignments, we conducted this analysis by classifying each student based on the three LLM support types they used most frequently among the three listed in Table~\ref{tab:llm_usage_categories}. Those who most frequently used usage types categorized as student-driven support were assigned to the \textit{student-driven support} group; those who most frequently used LLM usage types that are LLM-driven support were assigned to the \textit{LLM-driven support} group; and those who most frequently used LLM usage types that are comprehension support were assigned to the \textit{comprehension support} group. Based on this criterion, 23 students were included in the analysis: 14 were classified as LLM-driven support group and 7 as student-driven, with 2 students left unclassified due to equal frequencies across support types.

\begin{figure}[!t]
    \centering
    \includegraphics[width=1\linewidth]{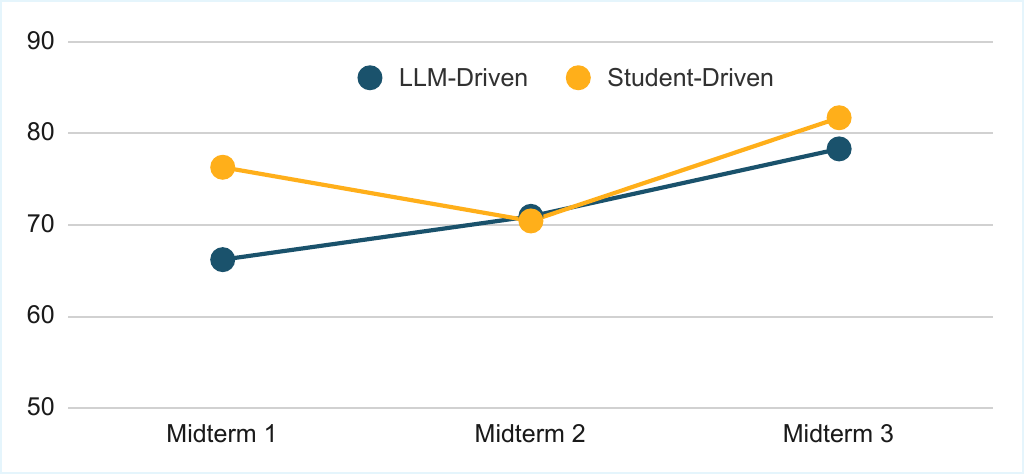}
    \caption{Comparisons of midterm exam scores between LLM-driven support group and Student-driven support group}
    \Description{The figure shows a line graph of midterm exam performance for LLM-driven support group and Student-driven support group. The x-axis represents the three midterms, and the y-axis represents average scores. Two lines trace the performance of each group, allowing comparison of trends across exams.}
    \label{fig:midterm_support_type}
\end{figure}

We then compared midterm exam scores between students whose LLM use was classified as LLM-driven and those classified as student-driven. Because the group sizes were highly imbalanced---similar to the comparison between the High-Reliance and Low-Reliance groups---we did not conduct statistical tests and instead observed the trends in the average midterm exam scores shown in Figure~\ref{fig:midterm_support_type}.
As the figure illustrates, students in the student-driven group scored higher on Midterm~1, but the performance of the two groups converged in subsequent midterms.
This pattern suggests that the type of support students sought---reflecting differences in how much effort they invested in contributing substantive work themselves---impacted their midterm performance early in the course, but that this effect may have diminished as students became more familiar with the course content and exam structure, similar to the comparison between LLM and No LLM groups.

\section{Discussion}
\label{sec:results_summary}
\vspace{5pt}
In this section, we summarize our results regarding students’ LLM usage and its impact on learning, as measured by midterm performance. 

We found that only a minority of students (23 out of 68; 34\%) used LLMs for the course assignments. Among these students, most (16 students) used LLMs lightly, relying on them for approximately 5–33\% of their submissions (Low-Reliance group). Their LLM use also tended to be concentrated earlier in the course, primarily before the first midterm. This pattern suggests that as these students became more confident with the course material and the exam structure, the perceived benefits of using LLMs may have decreased, reducing their incentive to rely on them.
The remaining seven students used LLMs more heavily (i.e., in more than 50\% of their submissions; High-Reliance group), either by using them consistently across assignments or by increasing their usage later in the course.

We observed a trend in which Low-Reliance students performed between 7\% and 8\% better on the three midterms than High-Reliance students, suggesting that the amount of LLM usage may be associated with differences in test performance in this context.

Our results also showed that students used LLMs in qualitatively different ways. Some forms of usage required substantial work and input from the student (student-driven support), while others relied more heavily on the LLM to generate content (LLM-driven support). We observed a trend in which students with a higher frequency of student-driven usage performed better on the midterms than students with a higher frequency of LLM-driven usage, particularly on the first midterm (a 10\% difference). This finding suggests that the type of LLM usage may be another factor influencing test performance in this context.

Overall, we found a statistically significant difference in performance on the first midterm between students who reported using LLMs on the assignments preceding this exam and students who reported no LLM use, with the latter group performing better with a large effect size. This difference decreased for the subsequent two midterms and, notably, the LLM-using group improved by approximately 10\% from the first midterm to the last.

Thus, although the type and amount of LLM use may have negatively affected student learning early on in the course, as the course progressed, most students appeared to calibrate their LLM use to better complement their learning processes. These findings provide encouraging evidence that the use of LLMs does not necessarily hinder learning, which is a major concern expressed within the educational community. Our results further suggest that many students are able to leverage LLMs constructively even without explicit guidance, given sufficient opportunity and time to experiment with their use. This offers initial evidence against the common concern that students will inevitably over-rely on LLMs in ways that undermine learning if left to their own devices. That said, our findings also indicate that providing guidance on effective LLM usage early in the course may help accelerate appropriate and productive adoption.

It should be noted, however, that in the course addressed in this paper, students could directly experience the impact of their LLM usage on performance through the midterms, as these assessments involved the same types of tasks as the assignments. In instructional settings where the connection between assignments and in-class exams is less explicit---and where the consequences of LLM use for exam performance are therefore less immediately observable---additional instructional scaffolding, feedback, or explicit guidance may be necessary to support effective and learning-oriented use of LLMs.

\section{Conclusion}
\vspace{5pt}
This work examined how students used LLMs in a research-oriented course, revealing diverse patterns of engagement and how these behaviors related to performance on an independently completed midterm exam. Our analyses showed meaningful variation in when students chose to use LLMs, the types of support they sought, and the amount of effort they contributed themselves, underscoring that LLM use is far from uniform.
Overall, the results offer an encouraging view: LLM use is not inherently detrimental, and several students demonstrated that, when used constructively, these tools can support---rather than hinder---learning.



Future work can build on this study in several important ways. First, although our analysis revealed clear patterns, the dataset---particularly within certain subgroups---was relatively small, meaning that some observations should be interpreted as trends rather than definitive effects. Replicating and extending these analyses with larger dataset would help establish the robustness and generalizability of our findings. Second, future work should also explore additional factors that may shape students' engagement with LLMs. For example, prior research highlights the role of AI literacy in influencing how students use these tools~\cite{kim2025students}, suggesting that students’ competencies, confidence, and prior experience with AI may meaningfully interact with their usage patterns. Other individual or contextual factors---such as study strategies or prior familiarity with research-based coursework---may also play a role and warrant further investigation.

Building on our findings summarized in Section~\ref{sec:results_summary}, we examined naturally occurring LLM usage behaviors and demonstrated how these categories relate to students' ability to perform in a new domain. By qualitatively analyzing the students' LLM-use reports to interpret their intentions and combining this with performance outcomes, we show that the manner in which students use LLMs---not merely the frequency of use---plays a meaningful role in learning.
In summary, our work advances understanding of whether and how students use LLMs, and how such usage affects learning, by examining LLM engagement in a cognitively demanding, interpretation-focused context and in an authentic course setting where usage patterns emerged organically. Together, these contributions represent one step toward a more comprehensive account of student–LLM interaction across different learning domains and underscore the importance of studying LLM use in ecologically valid environments.
\section{Acknowledgments}
This work was funded by the UBC Skylight Grant PM015111 and complies with UBC policies 81 and LR11 regarding ownership and usage of course material for pedagogical quality assurance. This work was also supported in part by the Institute for Computing, Information and Cognitive Systems (ICICS) at UBC.

%
\bibliographystyle{abbrv}
\bibliography{sigproc}  

@inproceedings{vadaparty2024cs1,
    title={{CS1-LLM: Integrating LLMs into CS1 instruction}},
    author = {Vadaparty, Annapurna and Zingaro, Daniel and Smith IV, David H. and Padala, Mounika and Alvarado, Christine and Gorson Benario, Jamie and Porter, Leo},
    year = {2024},
    publisher = {Association for Computing Machinery},
    booktitle = {Proceedings of the 2024 on Innovation and Technology in Computer Science Education V. 1},
    pages = {297–303},
}

@inproceedings{macneil2023experiences,
  title={{Experiences from Using Code Explanations Generated by Large Language Models in a Web Software Development E-Book}},
  author={MacNeil, Stephen and Tran, Andrew and Hellas, Arto and Kim, Joanne and Sarsa, Sami and Denny, Paul and Bernstein, Seth and Leinonen, Juho},
  booktitle={Proceedings of the 54th ACM Technical Symposium on Computer Science Education V. 1},
  pages={931--937},
  year={2023}
}

@inproceedings{rasnayaka2024empirical,
  title={{An Empirical Study on Usage and Perceptions of LLMs in a Software Engineering Project}},
  author={Rasnayaka, Sanka and Wang, Guanlin and Shariffdeen, Ridwan and Iyer, Ganesh Neelakanta},
  booktitle={Proceedings of the 1st International Workshop on Large Language Models for Code},
  pages={111--118},
  year={2024}
}

@inproceedings{lau2023ban,
  title={{From ``Ban It Till We Understand It'' to ``Resistance is Futile'': How University Programming Instructors Plan to Adapt as More Students Use AI Code Generation and Explanation Tools such as ChatGPT and GitHub Copilot}},
  author={Lau, Sam and Guo, Philip},
  booktitle={Proceedings of the 2023 ACM Conference on International Computing Education Research-Volume 1},
  pages={106--121},
  year={2023}
}

@article{mollick2023assigning,
  title={{Assigning AI: Seven Approaches for Students, with Prompts}},
  author={Mollick, Ethan and Mollick, Lilach},
  journal={arXiv preprint arXiv:2306.10052},
  year={2023}
}

@inproceedings{gao2024taxonomy,
  title={{A Taxonomy for Human-LLM Interaction Modes: An Initial Exploration}},
  author={Gao, Jie and Gebreegziabher, Simret Araya and Choo, Kenny Tsu Wei and Li, Toby Jia-Jun and Perrault, Simon Tangi and Malone, Thomas W},
  booktitle={Extended Abstracts of the CHI Conference on Human Factors in Computing Systems},
  pages={1--11},
  year={2024}
}

@article{santu2023teler,
  title={{TELeR: A General Taxonomy of LLM Prompts for Benchmarking Complex Tasks}},
  author={Santu, Shubhra Kanti Karmaker and Feng, Dongji},
  journal={arXiv preprint arXiv:2305.11430},
  year={2023}
}

@article{bull2023generative,
  title={{Generative Artificial Intelligence Assistants in Software Development Education: A Vision for Integrating Generative Artificial Intelligence into Educational Practice, Not Instinctively Defending Against It}},
  author={Bull, Christopher and Kharrufa, Ahmed},
  journal={IEEE Software},
  volume={41},
  number={2},
  pages={52--59},
  year={2023},
  publisher={IEEE}
}

@inproceedings{kazemitabaar2023novices,
  title={{How Novices Use LLM-based Code Generators to Solve CS1 Coding Tasks in a Self-Paced Learning Environment}},
  author={Kazemitabaar, Majeed and Hou, Xinying and Henley, Austin and Ericson, Barbara Jane and Weintrop, David and Grossman, Tovi},
  booktitle={Proceedings of the 23rd Koli calling international conference on computing education research},
  pages={1--12},
  year={2023}
}

@inproceedings{ghimire2024coding,
  title={{Coding with AI: How Are Tools Like ChatGPT Being Used by Students in Foundational Programming Courses}},
  author={Ghimire, Aashish and Edwards, John},
  booktitle={International Conference on Artificial Intelligence in Education},
  pages={259--267},
  year={2024},
  organization={Springer}
}

@inproceedings{brender2024s,
  title={{Who's Helping Who? When Students Use ChatGPT to Engage in Practice Lab Sessions}},
  author={Brender, J{\'e}r{\^o}me and El-Hamamsy, Laila and Mondada, Francesco and Bumbacher, Engin},
  booktitle={International Conference on Artificial Intelligence in Education},
  pages={235--249},
  year={2024},
  organization={Springer}
}

@article{kim2025students,
author = {Jinhee Kim and Seongryeong Yu and Sang-Soog Lee and Rita Detrick},
title = {Students’ prompt patterns and its effects in AI-assisted academic writing: Focusing on students’ level of AI literacy},
journal = {Journal of Research on Technology in Education},
volume = {58},
number = {3},
pages = {638--655},
year = {2026},
publisher = {Routledge},
doi = {10.1080/15391523.2025.2456043},
URL = {https://doi.org/10.1080/15391523.2025.2456043},
eprint = {https://doi.org/10.1080/15391523.2025.2456043}
}

@incollection{grande2024student,
  title={Student perspectives on using a large language model (llm) for an assignment on professional ethics},
  author={Grande, Virginia and Kiesler, Natalie and Francisco R, Mar{\'\i}a Andre{\'\i}na},
  booktitle={Proceedings of the 2024 on Innovation and Technology in Computer Science Education V. 1}, publisher = {Association for Computing Machinery},
  pages={478--484},
  year={2024}
}

@article{sawalha2024analyzing,
author = {Ghadeer Sawalha and Imran Taj and Abdulhadi Shoufan},
title = {{Analyzing student prompts and their effect on ChatGPT’s performance}},
journal = {Cogent Education},
volume = {11},
number = {1},
pages = {2397200},
year = {2024},
publisher = {Cogent OA},
doi = {10.1080/2331186X.2024.2397200},
}

@Inbook{ma2026examining,
author="Ma, Boxuan and Chen, Li and Konomi, Shin'ichi",
title="Examining Student-ChatGPT Interactions in Programming Education",
bookTitle="Teaching and Learning in the Generative Artificial Intelligence Age",
year="2025",
publisher="Springer Nature Switzerland",
address="Cham",
pages="97--114",
isbn="978-3-032-05817-1",
doi="10.1007/978-3-032-05817-1_5",
url="https://doi.org/10.1007/978-3-032-05817-1_5"
}

@article{sun2024investigating,
  title={Investigating students’ programming behaviors, interaction qualities and perceptions through prompt-based learning in ChatGPT},
  author={Sun, Dan and Boudouaia, Azzeddine and Yang, Junfeng and Xu, Jie},
  journal={Humanities and Social Sciences Communications},
  volume={11},
  number={1},
  pages={1--14},
  year={2024},
  publisher={Palgrave}
}

@article{schulhoff2024prompt,
  title={The prompt report: a systematic survey of prompt engineering techniques},
  author={Schulhoff, Sander and Ilie, Michael and Balepur, Nishant and Kahadze, Konstantine and Liu, Amanda and Si, Chenglei and Li, Yinheng and Gupta, Aayush and Han, HyoJung and Schulhoff, Sevien and others},
  journal={arXiv preprint arXiv:2406.06608},
  year={2024}
}

@article{giray2023prompt,
  title={Prompt engineering with ChatGPT: a guide for academic writers},
  author={Giray, Louie},
  journal={Annals of biomedical engineering},
  volume={51},
  number={12},
  pages={2629--2633},
  year={2023},
  publisher={Springer}
}

@article{jelson2025empirical,
  title={An empirical study to understand how students use ChatGPT for writing essays},
  author={Jelson, Andrew and Manesh, Daniel and Jang, Alice and Dunlap, Daniel and Kim, Young-Ho and Lee, Sang Won},
  journal={arXiv preprint arXiv:2501.10551},
  year={2025}
}

@article{flower1981cognitive,
  title={A Cognitive Process Theory of Writing},
  author={Flower, Linda and Hayes, John R.},
  journal={College Composition and Communication},
  volume={32},
  number={4},
  pages={365--387},
  year={1981},
  publisher={National Council of Teachers of English},
  doi={10.2307/356600}
}

@inproceedings{orozco2025emergent,
  title={An Emergent Bottom-Up Categorization of Students’ LLMs Usage in an Undergraduate Research Course},
  author={Orozco Vasquez, Ivan and Mahinpei, Romina and Elouazizi, Noureddine and Conati, Cristina},
  booktitle={International Conference on Artificial Intelligence in Education},
  pages={133--139},
  year={2025},
  organization={Springer}
}
%
\balancecolumns
\end{document}